# Hardware-Friendly Randomization: Enabling Random-Access and Minimal Wiring in FHE Accelerators with Low Total Cost

Ilan Rosenfeld, Noam Kleinburd, Hillel Chapman, Dror Reuven
Chain Reaction, Ltd.
{ilanr, noamk, hillelc, drorr}@chain-reaction.io

**Abstract**

*The Ring-Learning With Errors (RLWE) problem forms the backbone of highly efficient Fully Homomorphic Encryption (FHE) schemes. A significant component of the RLWE public key and ciphertext of the form $(\boldsymbol{b}, \boldsymbol{a})$ is the uniformly random polynomial $\boldsymbol{a} \in \mathcal{R}_Q$. While essential for security, the communication overhead of transmitting $\boldsymbol{a}$ from client to server, and inputting it into a hardware accelerator, can be substantial, especially for FHE accelerators aiming at high acceleration factors. A known technique in reducing this overhead generates $\boldsymbol{a}$ from a small seed on the client side via a deterministic process, transmits only the seed, and generates $\boldsymbol{a}$ on-the-fly within the accelerator. Challenges in the hardware implementation of an accelerator include wiring (density and power), compute area, compute power as well as flexibility in scheduling of on-the-fly generation instructions. This extended abstract proposes a concrete scheme and parameters wherein these practical challenges are addressed. We detail the benefits of our approach, which maintains the reduction in communication latency and memory footprint, while allowing parallel generation of uniformly distributed samples, relaxed wiring requirements, unrestricted random-access to RNS limbs, and results in an extremely low overhead on the client side (i.e. less than 3%) during the key generation process. The proposed scheme eliminates the need for thick metal layers for randomness distribution and prevents the power consumption of the PRNG subsystem from scaling prohibitively with the acceleration factor, potentially saving tens of Watts per accelerator chip in high-throughput configurations.*

## 1 Introduction and Motivation

### 1.1 Introduction

Fully Homomorphic Encryption (FHE) is gaining traction, with a growing number of industry applications relying on outsourced computation over encrypted data. The performance bottleneck has largely shifted from theoretical complexity to engineering challenges, particularly in minimizing latency and maximizing throughput via hardware acceleration. In the common client-server deployment model, the client prepares data and sends ciphertexts and public evaluation keys to the server, which hosts the FHE accelerator ASIC.

A standard RLWE[1]-based ciphertext is represented as a pair of polynomials $(\boldsymbol{b}, \boldsymbol{a})$, each in $\mathcal{R}_Q = \mathbb{Z}_Q[X]/(X^N + 1)$ where $\boldsymbol{a}$ is sampled uniformly in $\mathcal{R}_Q$, and $\boldsymbol{b} = \boldsymbol{m} - \boldsymbol{a} \cdot \boldsymbol{s} + \boldsymbol{e}$ (where $\boldsymbol{s}$ is the secret key polynomial, $\boldsymbol{m}$ the message to be encrypted, and $\boldsymbol{e}$ a randomly generated error polynomial). Homomorphic Encryption schemes allow for computation over data encrypted in such ciphertexts, up to a certain multiplicative depth, where depleted ciphertexts must either be decrypted by the client or undergo a bootstrapping[2], [3], [4] operation, which refreshes the multiplicative budget. For a typical security level (e.g. 128-bit) and supporting bootstrapping in schemes such as CKKS[5], BFV[6] and BGV[7], the ring dimension $N$ is high (e.g. $2^{16}$) and the modulus $Q$ may be large (e.g. $\sim 2^{1600}$[8]). Consequently, a single polynomial can reach a significant size (e.g. 13MB for the given examples).

The client produces even larger amounts of data in the form of public Key-Switching-Keys, which for the "Hybrid Key-Switching" method[9] consist of $\beta$ pairs of polynomials $(\boldsymbol{b}_i, \boldsymbol{a}_i)$ in $\mathcal{R}_Q$, thus summing to very large sizes (e.g. 130MB for the given examples and $\beta = 5$). Homomorphic applications may use many such keys amounting to GBs of data and thus may encumber the communications from client to server, and more crucially the communications within the server to load data into the accelerator.

The $\boldsymbol{a}_i$ polynomials are uniformly sampled from $\mathcal{R}_Q$ and are made public, thus a known system optimization[10], [11] generates these polynomials from a small seed via a deterministic process of PRNG, followed by rejection sampling, enabling the communication of just the seed instead of $\boldsymbol{a}_i$, reducing communication by half. The accelerator can then generate the $\boldsymbol{a}_i$ polynomials on-the-fly using dedicated hardware that performs an identical deterministic process.

### 1.2 Motivation and main contributions

ASIC accelerators tend to have a large amount of parallel computation resources (commonly modular arithmetic multipliers and adders), working at a high throughput, and requiring high demand for wiring, silicon area and power. Assigning a specific (perhaps central) area for the PRNG and then transmitting results to each parallel computation resource may incur high internal global wiring requirements due to the high rate at which data is produced and consumed. Our first motivation is to lower wiring cost, and our approach uses distributed PRNG generation, allowing the usage of local wiring between PRNG and computation resources.

Another detail in the implementation of seeded RLWE, is the serial nature of hash-based XOF used for the PRNG. Thus, to generate the $k^{th}$ output (for example a specific RNS limb of a Key-Switching-Key polynomial), outputs $0, 1, \dots, k-1$ must be produced beforehand. This limits flexibility in scheduling the homomorphic program, either constraining the program to use data in the order it is produced, or can be mitigated by storing the intermediate results, at the expense of memory resources. Our construction allows for random-access to the various RNS limbs, without the serial dependency.

The process of rejection sampling introduces some uncertainty as to the amount of PRNG required to generate enough accepted samples and thus might result in non-deterministic latency. This may complicate the compilation process as FHE programs can benefit greatly from static scheduling with deterministic latencies. Our construction addresses this issue, enabling a trade-off between computation on server side and client side, and the list of supported primes for the RNS. We demonstrate practical parameters for different trade-off possibilities.

## 2 Preliminaries

### 2.1 RLWE FHE Schemes

CKKS, BGV, and BFV are FHE schemes based on the Ring-Learning with Errors (RLWE) problem. Ciphertexts are typically pairs of polynomials in $\mathcal{R}_Q = \mathbb{Z}_Q[X]/(X^N + 1)$, and Key-Switching-Keys (utilizing the "Hybrid Key-Switching" method) consist of $\beta$ pairs of polynomials $(\boldsymbol{b}_i, \boldsymbol{a}_i) \in \mathcal{R}_Q^2$, in a modulus $Q$ and ring dimension $N$. Key-Switching-Keys may be very large due to large ring dimension (e.g. $2^{16}$), large modulus (e.g. $\sim 2^{1600}$) and $\beta$ (e.g. 5). Typical FHE programs involve the use of multiple different Key-Switching-Keys, and their total size can reach GBs, encumbering communications between client and server and specifically incur large loading time into FHE Accelerators.

### 2.2 Residue Number System (RNS)

The large moduli of a polynomial, $Q$, can be chosen to be a product of many smaller co-prime moduli $Q = \prod_{i=0}^{L-1} q_i$, enabling efficient modular arithmetic with smaller word-size moduli (e.g. using 32-bit word-size multipliers instead of 1600-bit ones) using the Chinese Remainder Theorem. Each residual polynomial (modulo a certain small modulus) is named "an RNS limb". We refer to $\{q_0, q_1, \dots, q_{L-1}\}$ as the RNS base of the polynomial. Uniformly distributed numbers within the large modulus $Q$ are expressed as a set of numbers uniformly distributed modulo each modulus $q_i$. In this work, we typically refer to uniform distribution over the unsigned range $[0, q_i)$ while the same principles work on the equivalent signed range $(-q_i/2, q_i/2)$.

### 2.3 Number Theoretic Transform (NTT)

Naive polynomial multiplication is a convolution, taking $O(N^2)$ operations. The Number Theoretic Transform transforms the regular polynomial representation (also known as Coefficient-mode) of a single limb into NTT representation (also known as Evaluation-mode), at the cost of $O(N \log N)$ operations. At this representation/domain polynomial multiplication is expressed as element-wise modular multiplication between the limbs. In the context of generating uniformly distributed Key-Switching-Key polynomials, these can be generated directly in the NTT domain, as their only usage is in polynomial multiplication during key-switching and the uniform distribution is maintained whether coefficients are sampled in regular or NTT-domain.

### 2.4 Seeded RLWE

As $\boldsymbol{a}_i$ are public polynomials, the client can generate them by a deterministic PRNG from a random seed, followed by rejection sampling, then sending only the seed to the server. In this way just half of the Key-Switching-Key may be communicated. The same, though not as common, can be applied to ciphertexts encrypted directly with the secret key.

### 2.5 Hash functions employed for PRNG

Hash functions have been used for the random generation of polynomials. FHE Libraries such as OpenFHE[12] and Lattigo[13] use the BLAKE2[14] function, while the MLWE CRYSTALS-Kyber[15] (selected by NIST for ML-KEM) uses SHAKE128[16] (part of the SHA-3 standard). All these instantiations use the Extendable Output Function (XOF) to generate an arbitrarily long hash output and thus operate serially; generating the $k^{th}$ output (which could be a specific limb), requires the prior computation of all preceding outputs $0, 1, \dots, k-1$. FHE Accelerator papers[11], [17] have also cited using KangarooTwelve[18] or Trivium[19] as algorithms for PRNG. We model these functions as random oracles for the purpose of this paper.

### 2.6 Support for a limited set of prime moduli

The basic limitations on supported prime moduli for the RNS base are that they be NTT-friendly (i.e. $q_i \bmod 2N \equiv 1$) and that each must fit within the hardware's word-size.

Optimization techniques may narrow the supported moduli set $S^{(q)}$ to such primes with specific characteristics that make modular reduction less costly. An example[20] for such characteristic is the use of primes of the form $q = 2^{a_1} \pm$ $2^{a_2} \ldots \pm 2^{a_h} + 1$, which make multiplication by $q$ (a common operation in modular reduction) consist of up to $h$ additions or subtraction. An equivalent formulation of the last expression is that the hamming weight (i.e. number of nonzero values) of the Non-Adjacent Form[21] (NAF) of $q$ is $\text{HW}_{\text{NAF}} = h + 1$. Limiting $\text{HW}_{\text{NAF}}$ to a certain maximum value narrows down from the entire set of NTT-friendly primes within the hardware's word-size.

The set of supported moduli, $S^{(q)}$, must be large enough to allow choosing a subset of $q_i$ such that $Q = \prod_{i=0}^{L-1} q_i$. The size of each modulus is also important due to its effect when performing Rescaling operations (in BFV and CKKS) or Modulus Switching (in BGV), where the actual value of $q_i$ is divided upon before rounding. Therefore, for a given homomorphic computation, with given precision requirements, $q_i$ must be at least of a given size, but not too large as the Rescaling/Modulus Switching operations consume this modulus.

Other modulus-consumption optimization techniques employed such as in [22], [23], may require support of some small-size moduli to utilize them in an efficient way.

### 2.7 Rejection sampling techniques

For a prime $q$, we are to sample a value of $m = \lceil \log q \rceil$ bits, distributed in the range $[0, q)$. We try to generate this distribution out of $n$ bits that are independently uniformly distributed (the result of a hash function for example).

For $q = 2^m - \varepsilon$ (with $\varepsilon \ll 2^{m-1}$) and $n = m$, the probability of rejection is $p_r = \frac{\varepsilon}{2^m} \ll 0.5$, while for $q = 2^{m-1} + \varepsilon$ (with $\varepsilon \ll 2^{m-1}$), and $n = m$, we have $p_r = \frac{2^{m-1} - \varepsilon}{2^m} \approx 0.5$ which is high.

Choosing $n = m + x$, where $x$ is a small number of extra bits, allows for lowering the rejection probability further. The first option of utilizing the extra bits without introducing bias to the distribution. This means that we can reject samples in $\left[\left\lfloor \frac{2^n}{q} \right\rfloor q, 2^n\right)$ and accept otherwise. The value should be further reduced to the range $[0, q)$. This improves the rejection probability to $p_r = \frac{2^n \bmod q}{2^n} < \frac{q}{2^{m+x}} < 2^{-x}$ (and in many cases much better than the bound). This implies that choosing $x = 2$ guarantees a rejection probability better than $1/4$, and $x = 5$ guarantees a rejection probability better than $1/32$.

A second option allows considering just the first $x$ bits (out of $n = m + x$) as candidates for MSB of an $m$-bit number and rejecting them immediately if the resulting number is to be in range $[q, 2^m)$, in which case there are still $m$ bits left for another attempt at sampling. If the result is not known to be in the rejection range, the remaining $m - x$ are filled from the remaining $m$. The probability of rejection decreases at a slower rate than the first option as a function of $x$, but the circuitry is simpler (no modular reduction required).

## 3 Optimized Seeded FHE Accelerator Scheme

### 3.1 Achieving low wiring requirements and random-access flexibility

FHE ASIC accelerators commonly have a large, parallel set of computation resources, comprising of mainly modular arithmetic multipliers and adders[11], [24], [25], [26]. The NTT-domain coefficients of the bespoke uniformly distributed polynomials are used as inputs to these computation resources. Since the computation resources are physically spread across the ASIC area, producing all these coefficients in a specific (perhaps central) area (Figure 1a) would require many, long wiring resources. These are costly in energy and availability (due to finite metal stack).

Our design (Figure 1b) divides a given limb of $N$ values (each of $w$ bits), into $n_{\text{seg}}$ segments of $\text{len} = N/n_{\text{seg}}$ values each. Each such segment is to be generated via PRNG independently from each other, allowing for a local PRNG engine be placed near a group of computation resources, with only local wiring. All existing hardware engines can work in parallel.

Each segment's hashing input is domain-separated[27] from other segments, by its $\text{id}_{\text{seg}} \in \{0, 1, \ldots, n_{\text{seg}} - 1\}$. To overcome the serial nature of the XOF in hash functions, and fulfill the motivation for random-access generation of limbs corresponding to our RNS base $\{q_0, q_1, \ldots q_{L-1}\}$ we introduce another domain-separation in the input, based on the modulus $q \in \{q_0, q_1, \ldots, q_{L-1}\}$ itself.

Algorithm 1 and Algorithm 2 are both meant for the client-side, generating a multi-residue polynomial in a way that can allow the FHE accelerator (on the server side) benefit from these characteristics.

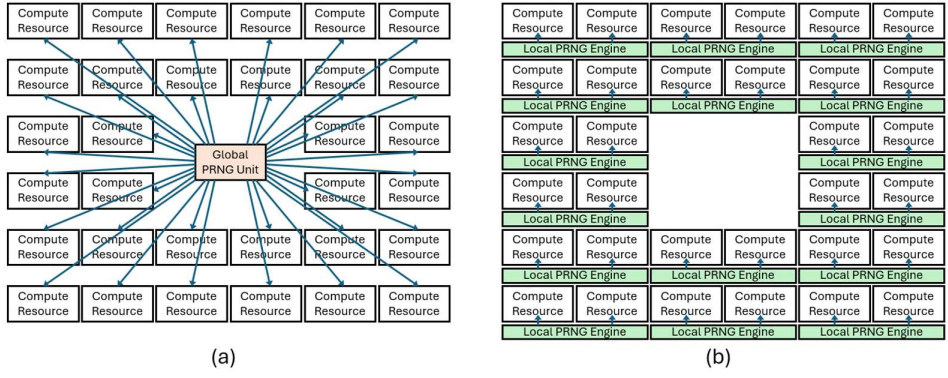

*Figure 1 - Central PRNG unit (a) vs Distributed PRNG unit (b)*

**Algorithm 1:** Output a uniformly distributed multi-residue polynomial (MRP) given an input `seed`, ring dimension $N$, RNS base `base` = $\{q_0, q_1, ... q_{L-1}\}$, number of segments $n_{seg}$, segment-generating function `GenSeg()`, and layout permutation matrix $P$. Each independent limb coefficient is of word-size $w$ bits.

```
    GenerateUniformlyDistributedMRP(seed, N, base, n_seg, P, w):
1       MRP ← zeroMRP(base, N)
2       len ← N/n_seg
3       for q ∈ base:
4         limb ← []
5         for id_seg from 0 to n_seg-1:
6           temp ← GenSeg(seed || q || id_seg, q, len, w)
7           if (length(temp) < len):
8               return error
9           else:
10              limb ← limb ○ temp
11        MRP[q] ← permute(limb, P)
12      return MRP
```

Algorithm 1 produces a multi-residue polynomial (MRP) by producing one limb corresponding to a modulus at a time. Each such RNS limb is divided into $n_{seg}$ segments, generated independently and domain-separated by both the modulus and the segment index $id_{seg}$. Note that || denotes **bit** concatenation and ○ denotes **list** appending. The `GenSeg()` function (detailed later in Algorithm 2) performs local PRNG and rejection sampling, aiming to produce a segment of $len = N/n_{seg}$ values uniformly distributed in the range [0, q). We account for the small probability that too few samples were accepted, and in this case return an error (lines 7-8) so that the algorithm may be re-run on the client side. Finally, in Algorithm 1 (line 11), we allow reordering of the coefficients within the limb to account for the fact that the physical hardware layout may not follow a standard sequential order; thus, NTT-domain coefficients that are physically close may not be consecutive in the logical sense. Therefore, a specific hardware configuration may require a permutation to reorder the coefficients from the device's native sequence back to the order the client expects ($b_i$ and $a_i$ must have matching order).

**Algorithm 2:** Generate a segment of uniformly distributed values in the range [0, $q$) of length `len` and word-size $w$, from a hash input `input`.

```
    GenSeg(input, q, len, w):
1       temp ← hash(input)
2       r ← length_in_bits(temp)
3       b ← 2^ceiling(log2(q))
4       t ← floor(r/w)
5       thresh ← floor(2**w/q)*q
6       seg ← []
7       for i from 0 to t-1:
8         sample ← temp[i*w:(i+1)*w]
9         if uint(sample) < thresh:
10          seg ← seg ○ sample
11        if length(seg) == len:
12          return seg
13      return seg
```

GenSeg(input, q, len, w), as detailed in Algorithm 2 works by hashing the input string, using a hash function that outputs a fixed number of bits $r$. For simplicity, we divide this output into $t = \left\lfloor \frac{r}{w} \right\rfloor$ words of $w$ bits each. For a given modulus $q$, a $w$-bit sample is considered as an unsigned integer and is accepted (lines 9-10) only if it is within the range $\left[0, \left\lfloor \frac{2^w}{q} \right\rfloor \cdot \frac{q}{2}\right)$. If enough samples have been collected (line 11), the segment is ready and the rest of the candidate samples are discarded.

Note that the sample is not explicitly reduced to the range $[0, q)$. We assume the downstream arithmetic unit (e.g., a modular multiplier) handles unreduced $w$-bit inputs without affecting the result.

### 3.2 Simplifying the design for deterministic scheduling and low hardware footprint

As seen in Algorithm 2, the hash(input) function produces a finite (pre-determined) number of bits $r$, that are split into $t = \left\lfloor \frac{r}{w} \right\rfloor$ words. This is followed by attempting to find len out of $t = \left\lfloor \frac{r}{w} \right\rfloor$ words that are in the required range. Increasing the value of $r$ increases hashing cost, and the rejection sampling cost (more comparators and multiplexer units), but also decreases the chance of failure. Increasing len makes more efficient use of the $r$ bits produced but increases the chance of failure to produce the number of required samples.

The probability of a single sample being rejected is $p_r = \frac{2^w \bmod q}{2^w}$, while the probability of GenSeg() successfully generating a segment of length $len$ is $p_{seg} = \sum_{i=\text{len}}^{t} \binom{t}{i} p_r^{t-i} (1 - p_r)^i$. The probability of successfully generating a whole limb is $p_{\text{limb}} = p_{seg}^{n_{seg}}$.

The probability of successfully generating a whole MRP is $p_{\text{MRP}} = \prod_{q_i \in \text{base}} p_{\text{limb},q_i} \geq p_{\text{limb, worst}}^{|\text{base}|}$ where $p_{\text{limb},q_i}$ is the probability of generating a whole limb for $q_i$, and $p_{\text{limb, worst}} = \min(p_{\text{limb},q_i} | q_i \in \text{base})$.

Each failure in producing an MRP results in an extra attempt with a different seed on the client-side, while the server accelerator may be guaranteed to always get parameters that will result in generation success. The probability of MRP generation failure should be low enough so that the extra client effort in generating the polynomial is minor, and such that the number of possible valid seeds is large enough (from the security perspective).

Setting the probability of MRP generation failure at a certain bound, e.g. $1 - p_{MRP} \leq 3\%$ lowers the number of possible seeds from $2^{\text{seedlength}}$ to $2^{\text{seedlength}} \cdot 0.97 = 2^{\text{seedlength}+\log 0.97}$, which can be made large enough for a large enough seed length.

We note that for typical ring dimensions (e.g. $2^{16}$) there are more than enough FHE-friendly primes of various sizes to produce RNS bases for typical ciphertext modulus size (e.g. $\sim 2^{1600}$). Therefore, we propose limiting the supported primes moduli to those such that $p_r$ is low enough. In section 4, we shall detail an example choice.

### 3.3 Benefits gained by the proposed construction

If a multi-residue polynomial was generated by the client using the above construction (Algorithms 1, 2 and extra attempts if necessary), this enables the server-side accelerator to benefit in the following ways:

**Distributed Generation**: each local engine performs only GenSeg(seed || q || id_seg, q, len, w). It is guaranteed to generate enough samples. The difference between engines is just the input bit string, where **seed** is common for all engines and all limbs, q is common to all engines when generating a specific limb, and id_seg is unique for each engine.

**Random Access:** if a program running on the accelerator is required to access a certain limb of a multi-residue polynomial, it can do so directly, by invoking the generation instruction with appropriate seed and modulus q. An example can be a computation of the form: $\sum_{i=0}^{\beta-1} d_i \cdot \text{KSK}_i$ which is typical during key-switching, where the $\text{KSK}_i$ are polynomials to be generated on-the-fly. Computation of a sum over one of the RNS limbs of these polynomials, such as corresponding to a modulus $q$, may be done directly as $\sum_{i=0}^{\beta-1} d_i[q] \cdot \text{KSK}_i[q]$ without the need to generate and store any of the other limbs in register file or memory.

**Scheduling friendliness:** By delegating seed validation to the client, the accelerator enjoys a deterministic execution profile, eliminating the need for complex flow control or back-pressure signals in the randomness distribution network. No stalls or bubbles will be introduced during run-time.

### 3.4 Gain in power and wiring implementation

For an accelerator chip with $R$ parallel lanes of computation units (e.g. modular multipliers), accepting inputs of word-size $w$, and working at frequency $f$, the maximum throughput that may be required of the uniformly-distributed-polynomial generation unit is $TP = \gamma \cdot R \cdot w \cdot f$, where $0 < \gamma \leq 1$ is a factor accounting for the fact that the $R$ parallel lanes are not fully occupied with multiplication by $a_i$ polynomials.

For a square chip of side $d = 15$mm, with $R = 16384$ multipliers of 32-bit word-size, laid-out uniformly over the chip area, and working at $f = 1$GHz, and assuming $\gamma = \frac{1}{8}$, we have $TP = 65.5$ Tbps. If implementing by a central unit, data travels an average Manhattan distance equal to $\frac{d}{2} = 7.5$mm, at a bandwidth density of $2.2 \frac{\text{Tbps}}{\text{mm}}$ in each axis assuming the whole $d = 15$mm cross-section can be used for wiring (density is much higher than at the central unit's edges). The wiring is for long reach (in on-chip scale), and at an energy cost of $E = 40 \frac{\text{fJ}}{\text{bit}\cdot\text{mm}}$ (as in [28]) this results in a power consumption of $P = 19.7W$. This power consumption and bandwidth density can be much aggravated at higher required throughputs.

The power cost associated with data travel is essentially eliminated with the distributed implementation, with the PRNG local engine being adjacent to the computation units. The maximum bandwidth density is not changed in the distributed implementation, however since these are very short-reach wires, very fine pitch conductors can be used, posing a much-reduced wiring challenge. Usage of thick metal layers for the task is eliminated.

## 4 Example parameters and trade-offs

### 4.1 Main area/power vs supported moduli set trade-off

In our example hardware, we support generation via SHAKE128 and KangarooTwelve, both are hash functions of the Keccak[29] family, with output length fixed by us at $r = 1344$ bits (avoiding multiple "squeezing" steps of the sponge function). Our design uses a $w = 32$ bit word-size, thus the hash function produces $t = \lfloor \frac{r}{w} \rfloor = 42$ words.

We limit our moduli set $S^{(q)}$ to include only NTT-friendly primes with ones satisfying $HW_{NAF} \leq 5$ for compact modular reduction hardware. In a typical 128-bit security application with $N = 2^{16}$, we assume a maximum modulus of $Q \approx 2^{1747}$ so we can safely assume an RNS base covering this modulus can be found with a few tens of moduli.

Note that for a fixed throughput of uniformly distributed polynomial generation, the value of len is <u>inversely proportional</u> to the <u>area and power</u> that the entire hash generation hardware takes within the accelerator chip. For values of len $\in \{4,8,16,32\}$, we calculate $p_{r,\max}$ such that the overall probability of an MRP rejection $1 - p_{\text{MRP}}$ is upper bounded by an arbitrarily value of 3% (thus client-side overhead is not very significant), and explore the statistics of available modulus sizes. Table 1 lists the results (both moduli size histogram and $p_{r,\max}$ for a given len).

*Table 1 - Statistics of moduli sizes in the supported set vs the maximum rejection probability per sample, with* len *fixed to a power-of-two, and the upper bound on probability of MRP rejection is less than 3%*

| | | Histogram of $\{\lceil \log q_i \rceil \mid q_i \in S^{(q)}\}$ | | | | | | | | | | | | | |
|---|---|---|---|---|---|---|---|---|---|---|---|---|---|---|---|
| $p_{r,\max}$ | $\lvert S^{(q)} \rvert$ | 20 | 21 | 22 | 23 | 24 | 25 | 26 | 27 | 28 | 29 | 30 | 31 | 32 | len | $(1 - p_{\text{MRP}}) \leq$ |
| 0.03655 | 277 | 2 | 1 | 1 | 8 | 15 | 18 | 26 | 51 | 39 | 37 | 20 | 27 | 32 | 32 | 3.00% |
| 0.25305 | 526 | 2 | 1 | 1 | 8 | 15 | 18 | 26 | 52 | 57 | 87 | 114 | 68 | 77 | 16 | 3.00% |
| 0.42359 | 562 | 2 | 1 | 1 | 8 | 15 | 18 | 26 | 52 | 57 | 87 | 115 | 98 | 82 | 8 | 3.00% |
| 0.5 | 625 | 2 | 1 | 1 | 8 | 15 | 18 | 26 | 52 | 57 | 87 | 115 | 161 | 82 | 4 | 0.29% |

Sampling with a word size of 32-bit maintains a very low probability of rejection sampling $p_r$ for primes of 27 bits or less and thus reducing len improves only the number of larger primes supported for the same given MRP rejection probability.

### 4.2 Complimentary Example parameters

The seed is chosen to be of length 288 bit – allowing, for example, comfortably having a 256-bit seed part that is common to all $\beta$ polynomials within a single key-switching-key, plus a 32-bit seed part that is unique per polynomial. For a given moduli base, with $(1 - p_{\text{MRP}}) \leq 3\%$ the number of valid seeds is at least $2^{288}(1 - 0.03) > 2^{287.95}$, which is large enough for security.

Other implementation details include:

- The hash input string contains $q$ – a 32-bit integer in the range 0 to $2^{32} - 1$.
- $\text{id}_{\text{seg}}$ is a 16-bit integer in the range 0 to 2047 (for $N = 2^{16}$).
- This totals in 336 bits (i.e. input = $\text{seed}_{288} || q_{32} || \text{id}_{16}$) as the input string size into the hash function. Inputs are further padded to a minimum size within the hash function according to the standard.

## 5  Conclusion and further work

This work has presented a construction for generating uniformly distributed polynomials on the client-side in a way that enables significant architectural efficiencies when performing homomorphic computations on a server-side accelerator. As acceleration speed increases, concerns of internal wiring become more prominent, and the construction eases requirements in this respect, notably reducing the power consumption and thick-metal layer usage significantly. The construction allows random-access to specific limbs of uniformly distributed polynomials, removing the constraint of serially generating all limbs until the required ones, or alternatively the memory requirement of generating all beforehand. A trade-off between the size of the supported moduli set and the required area and power has been demonstrated, with a small effect on the client-side required computation (which anyway happens once during the key-switching-key generation stage).

Wide adoption of this scheme (or one inspired by its principles) by FHE Accelerator vendors and FHE libraries/compilers would foster a more efficient ecosystem, allowing large encrypted applications to scale with lower energy and hardware overhead.

Future work may examine finer-grained rejection sampling, utilizing all the bits generated by a hash function (whereas in our construction some bits are discarded). Another aspect that may be examined, is random-access at other-than-limb dimension, i.e. the sample generation mechanism may be required to generate several different RNS limbs of the same coefficient index. Other hash primitives may be considered as well.